\documentclass[twocolumn]{aastex62}
\usepackage{amsmath,amstext}
\usepackage[T1]{fontenc}
\usepackage{apjfonts} 
\usepackage[figure,figure*]{hypcap}

\graphicspath{{./}{figures/}}

\received{April 26, 2019}
\revised{June 29, 2019}
\accepted{July 2, 2019}

\newcommand{\Msun}{\ensuremath{\mathrm{M}_\odot}}

\newcommand{\kmps}{\ensuremath{\mathrm{km~s^{-1}}}}

\newcommand{\Ha}{\ensuremath{\mathrm{H}\alpha}}
\newcommand{\Hb}{\ensuremath{\mathrm{H}\beta}}
\newcommand{\Hg}{\ensuremath{\mathrm{H}\gamma}}
\newcommand{\Hd}{\ensuremath{\mathrm{H}\delta}}

\shorttitle{HSC16aayt}
\shortauthors{T. J. Moriya et al.}

\begin{document}

\title{HSC16aayt: Slowly evolving interacting transient rising for more than 100 days}

\correspondingauthor{Takashi J. Moriya}
\email{takashi.moriya@nao.ac.jp}

\author[0000-0003-1169-1954]{Takashi J. Moriya}
\affiliation{National Astronomical Observatory of Japan, National Institutes of Natural Sciences, 2-21-1 Osawa, Mitaka, Tokyo 181-8588, Japan}
\author{Masaomi Tanaka}
\affiliation{Astronomical Institute, Tohoku University, 6-3 Aramaki Aza-Aoba, Aoba, Sendai, Miyagi 980-8578, Japan}
\author{Tomoki Morokuma}
\affiliation{Institute of Astronomy, Graduate School of Science, The University of Tokyo, 2-21-1 Osawa, Mitaka, Tokyo 181-0015, Japan}
\author{Yen-Chen Pan}
\affiliation{National Astronomical Observatory of Japan, National Institutes of Natural Sciences, 2-21-1 Osawa, Mitaka, Tokyo 181-8588, Japan}
\author{Robert M. Quimby}
\affiliation{Department of Astronomy / Mount Laguna Observatory, San Diego State University, 5500 Campanile Drive, San Diego, CA, 92812-1221, USA}
\affiliation{Kavli Institute for the Physics and Mathematics of the Universe (WPI), The University of Tokyo Institutes for Advanced Study, The University of Tokyo, 5-1-5 Kashiwanoha, Kashiwa, Chiba 277-8583, Japan}

\author{Ji-an Jiang}
\affiliation{Institute of Astronomy, Graduate School of Science, The University of Tokyo, 2-21-1 Osawa, Mitaka, Tokyo 181-0015, Japan}
\author{Kojiro Kawana}
\affiliation{Department of Physics, Graduate School of Science, The University of Tokyo, 7-3-1 Hongo, Bunkyo, Tokyo 113-0033, Japan}
\author{Keiichi Maeda}
\affiliation{Department of Astronomy, Kyoto University, Kitashirakawa-Oiwake-cho, Sakyo-ku, Kyoto 606-8502, Japan}
\affiliation{Kavli Institute for the Physics and Mathematics of the Universe (WPI), The University of Tokyo Institutes for Advanced Study, The University of Tokyo, 5-1-5 Kashiwanoha, Kashiwa, Chiba 277-8583, Japan}
\author{Ken'ichi Nomoto}
\affiliation{Kavli Institute for the Physics and Mathematics of the Universe (WPI), The University of Tokyo Institutes for Advanced Study, The University of Tokyo, 5-1-5 Kashiwanoha, Kashiwa, Chiba 277-8583, Japan}
\author{Nao Suzuki}
\affiliation{Kavli Institute for the Physics and Mathematics of the Universe (WPI), The University of Tokyo Institutes for Advanced Study, The University of Tokyo, 5-1-5 Kashiwanoha, Kashiwa, Chiba 277-8583, Japan}
\author{Ichiro Takahashi}
\affiliation{Kavli Institute for the Physics and Mathematics of the Universe (WPI), The University of Tokyo Institutes for Advanced Study, The University of Tokyo, 5-1-5 Kashiwanoha, Kashiwa, Chiba 277-8583, Japan}
\author{Masayuki Tanaka}
\affiliation{National Astronomical Observatory of Japan, National Institutes of Natural Sciences, 2-21-1 Osawa, Mitaka, Tokyo 181-8588, Japan}
\author{Nozomu Tominaga}
\affiliation{Department of Physics, Faculty of Science and Engineering, Konan University, 8-9-1 Okamoto, Kobe, Hyogo 658-8501, Japan}
\affiliation{Kavli Institute for the Physics and Mathematics of the Universe (WPI), The University of Tokyo Institutes for Advanced Study, The University of Tokyo, 5-1-5 Kashiwanoha, Kashiwa, Chiba 277-8583, Japan}
\author{Masaki Yamaguchi}
\affiliation{Department of Physics, Faculty of Science and Engineering, Konan University, 8-9-1 Okamoto, Kobe, Hyogo 658-8501, Japan}
\author{Naoki Yasuda}
\affiliation{Kavli Institute for the Physics and Mathematics of the Universe (WPI), The University of Tokyo Institutes for Advanced Study, The University of Tokyo, 5-1-5 Kashiwanoha, Kashiwa, Chiba 277-8583, Japan}

\author{Jeff Cooke}
\affiliation{Centre for Astrophysics \& Supercomputing, Swinburne University of Technology, Hawthorn, VIC 3122, Australia}
\affiliation{ARC Centre of Excellence for All-Sky Astrophysics (CAASTRO)}
\author{Chris Curtin}
\affiliation{Centre for Astrophysics \& Supercomputing, Swinburne University of Technology, Hawthorn, VIC 3122, Australia}
\affiliation{ARC Centre of Excellence for All-Sky Astrophysics (CAASTRO)}
\author[0000-0002-1296-6887]{Llu\'is Galbany}
\affiliation{PITT PACC, Department of Physics and Astronomy, University of Pittsburgh, Pittsburgh, PA 15260, USA}
\author{Santiago Gonz\'alez-Gait\'an}
\affiliation{CENTRA, Instituto Superior T\'ecnico, Universidade de Lisboa, Portugal}
\author{Chien-Hsiu Lee}
\affiliation{National Optical Astronomy Observatory, 950 North Cherry Avenue, Tucson, AZ 85719, USA}
\author{Giuliano Pignata}
\affiliation{Departamento de Ciencias F\'isicas, Universidad Andres Bello, Avda. Rep\'ublica 252, Santiago, 8320000, Chile}
\affiliation{Millennium Institute of Astrophysics (MAS), Nuncio Monse\~nor S\'otero Sanz 100, Providencia, Santiago, Chile}
\author{Tyler Pritchard}
\affiliation{Center for Cosmology and Particle Physics, New York University, 726 Broadway, New York, NY 10004, USA}


\begin{abstract}
We report our observations of HSC16aayt (SN~2016jiu), which was discovered by the Subaru/Hyper Suprime-Cam (HSC) transient survey conducted as part of Subaru Strategic Program (SSP). It shows very slow photometric evolution and its rise time is more than 100~days. The optical magnitude change in 400~days remains within 0.6 mag. Spectra of HSC16aayt show a strong narrow emission line and we classify it as a Type~IIn supernova. The redshift of HSC16aayt is $0.6814\pm0.0002$ from the spectra.
Its host galaxy center is at 5 kpc from the supernova location and HSC16aayt might be another example of isolated Type IIn supernovae, although the possible existence of underlying star forming activity of the host galaxy at the supernova location is not excluded.
\end{abstract}


\section{Introduction}\label{sec:introduction}

\begin{figure*}
 \begin{center}
  \includegraphics[width=\columnwidth]{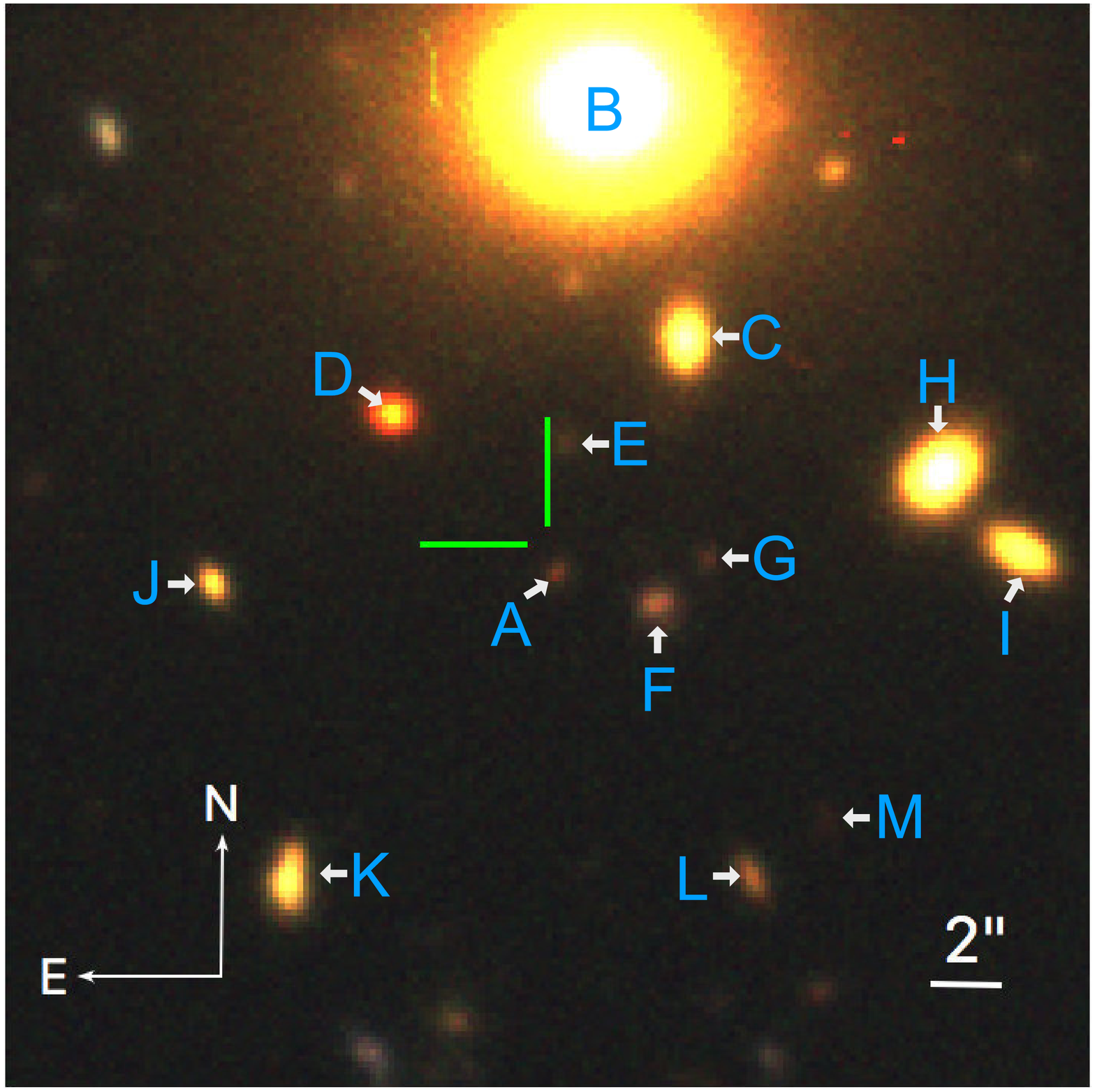}
  \includegraphics[width=\columnwidth]{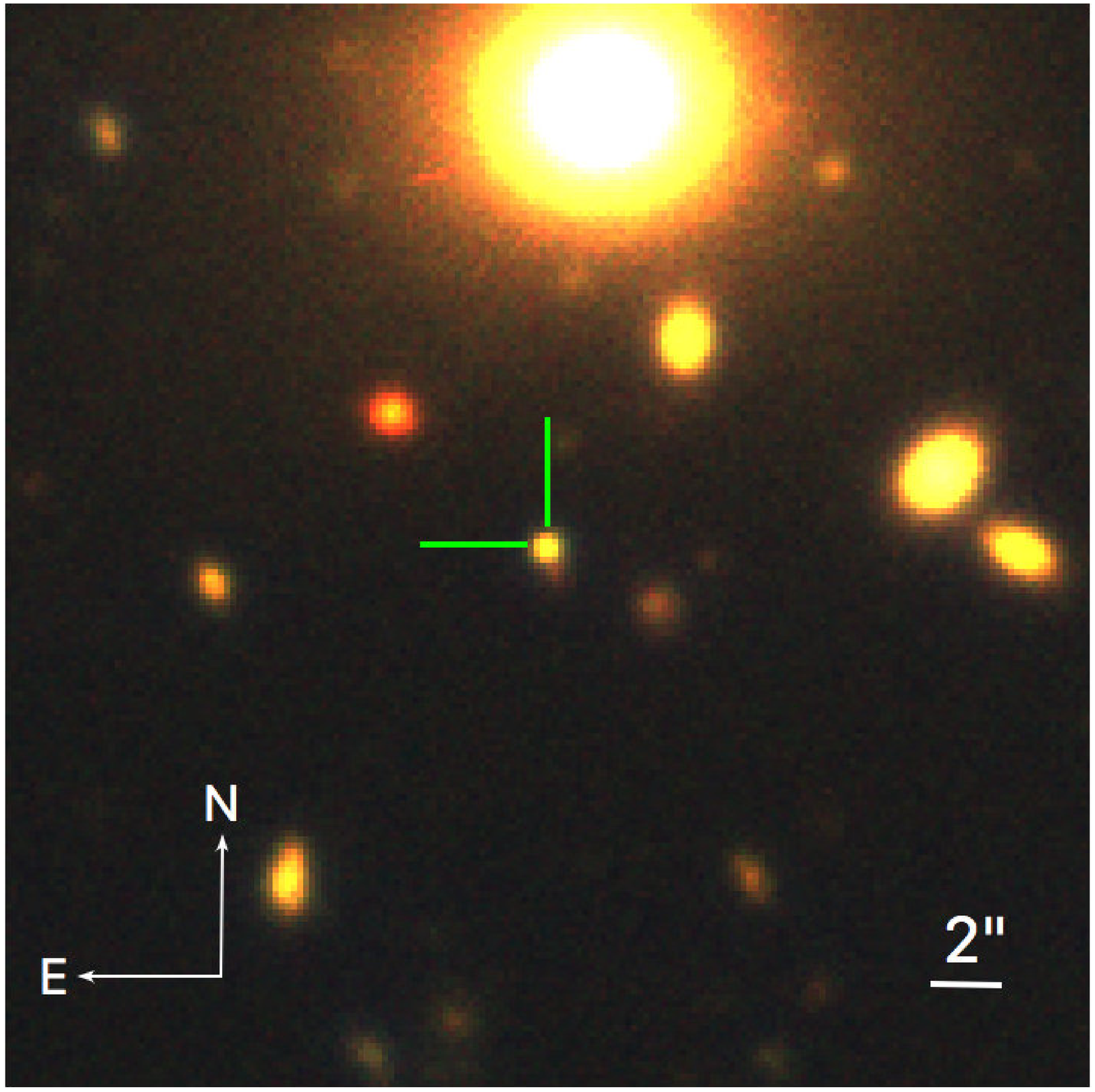}  
 \end{center}
\caption{
The $g$, $r$, and $i$ composite HSC images before (left) and after (right) the explosion of HSC16aayt. The image size is 30" x 30" and the SN location is at the center.
Galaxies near HSC16aayt are labeled in the left panel. The apparent nearest galaxy (A) has the photometric redshift estimate of $z=1.45^{+0.06}_{-0.06}$ from the deblended photometry.  The photometric redshifts of the other apparently nearby galaxies are $0.24^{+0.02}_{-0.03}$ (B), $0.50^{+0.02}_{-0.02}$ (C), $0.72^{+0.04}_{-0.04}$ (D), $0.76^{+0.16}_{-0.20}$ (E), $1.39^{+0.02}_{-0.03}$ (F), $1.41^{+0.48}_{-0.08}$ (G), $0.46^{+0.02}_{-0.02}$ (H), $0.49^{+0.01}_{-0.01}$ (I), $2.04^{+0.04}_{-0.04}$ (J), $0.50^{+0.03}_{-0.04}$ (K), $0.58^{+0.04}_{-0.04}$ (L), and $0.55^{+0.27}_{-0.16}$ (M). These photometric redshifts are estimated by the deblended photometry.
The galaxy B has the spectroscopic redshift of $z=0.2505$ obtained by SDSS.
}\label{fig:face}
\end{figure*}

Type~IIn supernovae (SNe, \citealt{schlegel1990iin}) are a subclass of hydrogen-rich SNe showing narrow emission lines in their spectra. The narrow lines are believed to originate from the interaction between SN ejecta and dense circumstellar media (CSM) formed by the progenitors before their explosion \citep[e.g.,][]{1993ApJ...414L.101C,chugai1994iin,fransson2002sn95n}. The mass-loss rates of the SN~IIn progenitors are estimated to be more than $\sim 10^{-4}~\Msun~\mathrm{yr^{-1}}$ \citep[e.g.,][]{kiewe2012iin,taddia2013iin,moriya2014iinhistory,fox2013iin,ofek2014ApJ...788..154Ointpeak} which is beyond the mass-loss rates of ordinary stars \citep[e.g.,][]{smith2014reveiw}. The progenitors of the most luminous SNe~IIn, which are also called superluminous SNe (SLSNe), are estimated to have the mass-loss rates of more than $\sim 0.1~\Msun~\mathrm{yr^{-1}}$ shortly before the explosions \citep[e.g.,][]{chevalier2011irwin,moriya2013sn06gy,chatzopoulos2013chi2,dessart2015slsniin}.

Some massive stars such as luminous blue variables (LBVs, \citealt{humphreys1994lbv}) are known to experience huge mass-loss rates exceeding $\sim 10^{-3}~\Msun~\mathrm{yr^{-1}}$ and they are suggested to be progenitors of SNe~IIn \citep[e.g.,][]{smith2010sn06gy,smith2011sn10jl,justham2014lbv}. Indeed, \citet{gal-yam2009sn05glprog} confirm the disappearance of a very massive star that is likely an LBV after the explosion of SN~IIn 2005gl. However, the progenitor of SN~IIn 2008S is found to be a relatively low mass star ($\simeq 10~\Msun$, \citealt{prieto2008sn08s,botticella2009sn08s,adams2016gone,eldridge2019ngc6946}). Furthermore, the explosion sites of SNe~IIn are not necessarily in strong favor of very massive star progenitors and they also indicate that various progenitors exist in SNe~IIn \citep{anderson2012env,habergham2014intenv,taddia2015iinmetal,galbany2018pisco}. It is, therefore, very likely that there are several stellar evolutionary paths leading to SNe~IIn \citep[cf.][]{yoon2010cantiello,dwarkadas2011iinprog,justham2014lbv,mackey2014pico,woosley2015lowmass,woosley2017ppisn}.

The strong interaction signatures in SNe~IIn prevent us from directly identifying their inner explosion properties. To form SNe~IIn, the dense CSM need to exist but the explosions inside do not need to be SN explosions \citep[e.g.,][]{woosley2007sn06gy,dessart2009sn94w}. In fact, some transients having SN~IIn spectra are known to be ``SN impostors'' \citep[e.g.,][]{vandyk2000imposter}. Their central explosions are not SN explosions and are related to stellar eruptive events like LBV eruptions \citep[e.g.,][]{vink2012lbv}. In some SNe~IIn, discussion on the true nature of the central explosions has not yet been converged. A notorious expample is SN~2009ip \citep{prieto2013sn09ip,fraser2013sn09ip,fraser2015sn09ip,pastorello2013sn09ip,smith2013sn09ip,smith2014sn09ip,mauerhan2013sn09ip,mauerhan2014sn09ip,ofek2013sn09ip,kashi2013sn09ip,soker2013sn09ip,tsebrenko2013sn09ip,ouyed2013sn09ip,margutti2014sn09ip,graham2014sn09ip,graham2017sn09ip,levesque2014sn09ip,moriya2015sn09ip,martin2015sn09ip,reilly2017sn09ip}. No clear conclusion on the nature of its central explosion that occurred in 2012 has yet been made.

One characteristic of SNe~IIn is their rich diversity in light curves (LCs, e.g., \citealt{kiewe2012iin,stritzinger2012iin,taddia2013iin}). The major luminosity source of SNe~IIn is the interaction between the ejecta and dense CSM and the mass-loss diversity in the progenitors makes the LC diversity \citep[e.g.,][]{moriya2013iin,moriya2014iinhistory}. Some SNe~IIn rise in about 10~days \citep[e.g.,][]{rest2011sn03ma,zhang2012sn2010jl}, while others take about 100~days to reach its peak luminosity \citep[e.g.,][]{elias-rosa2018snhunt151}. SN~IIn LCs sometimes even show bumpy structure \citep{nyholm2017bump}. The longest reported rise time of SNe~IIn is around 400~days found in SN~2008iy \citep{miller2010sn2008iy}, although SNe~IIn with such a long rise are quite rare. In this paper, we report the observations of HSC16aayt, which is a SN~IIn having extremely slow LC evolution as found in SN~2008iy. It was discovered by Subaru/Hyper Suprime-Cam (HSC, \citealt{miyazaki2018hsc,komiya2018hsc,furusawa2018hsc,kawanomoto2018hsc}) during the transient survey conducted as part of Subaru Strategic Program (SSP, \citealt{aihara2018hscssp}). 

The rest of this paper is organized as follows. First, we summarize our observations of HSC16aayt in Section~\ref{sec:observations}. We discuss the redshift and host galaxy of HSC16aayt in Section~\ref{sec:redshift}. We show the LC and spectral properties of HSC16aayt in Sections~\ref{sec:lightcurve} and \ref{sec:spectra}, respectively.
We discuss our results in Section~\ref{sec:discussion}.
We summarize this paper in Section~\ref{sec:summary}. We assume the standard cosmology with $H_0=70~\mathrm{km~s^{-1}~Mpc^{-1}}$, $\Omega_M = 0.3$, and $\Omega_\Lambda = 0.7$ in this paper.

\section{Observations}\label{sec:observations}
\subsection{Discovery}
HSC16aayt (SN~2016jiu) was detected throughout the HSC SSP transient survey in COSMOS that was performed from Nov. 2016 to Apr. 2017 \citep{yasuda2019ssp}. The first survey data were obtained on 23 Nov. 2016 (UT is used throughout this paper) and it was discovered on that day at (RA, Dec) = (10:02:05.57, +02:57:58.3). It was not in our reference images that were created from the data taken between Mar. 2014 and Apr. 2016. Fig.~\ref{fig:face} shows the reference and SN images. The Galactic extinction towards the COSMOS field is negligible and we do not find any signatures of strong host extinction. Therefore, we apply no extinction corrections to our observations.

One of the science objectives of the HSC SSP transient survey is the high-redshift SLSN survey (Subaru HIgh-Z sUpernova CAmpaign, SHIZUCA, \citealt{moriya2018hscslsn,curtin2018hscslsn}). The LC of HSC16aayt evolved very slowly and the apparent nearest galaxy on the sky located at 0.7\arcsec away to the south west (A in Fig.~\ref{fig:face}) had the photometric redshift of $1.45^{+0.06}_{-0.06}$, which is estimated by the \texttt{MIZUKI} code \citep{tanaka2015mizuki}. These two facts led us to speculate that HSC16aayt might be a high-redshift SLSN and we triggered the spectroscopic follow-up observations. The spectroscopic observations revealed that HSC16aayt is actually at a lower redshift as we present below.

\startlongtable
\begin{deluxetable}{cccc}
\tablecaption{Photometric data of HSC16aayt. \label{tab:photometory}}
\tablecolumns{4}
\tablehead{
\colhead{Band} & \colhead{MJD} &
\colhead{AB mag} & \colhead{$1\sigma$ error}  
}
\startdata
$g$&57755.62 & 23.587 &  0.013 \\
&57778.44 & 23.616 &  0.015 \\
&57785.38 & 23.369 &  0.033 \\
&57807.37 & 23.644 &  0.016 \\
&57834.31 & 23.650 &  0.012 \\
&57841.29 & 23.626 &  0.019 \\
&57869.33 & 23.588 &  0.023 \\
\hline
$r$&57720.59 & 23.201 &  0.023 \\
&57747.53 & 23.160 &  0.019 \\
&57776.40 & 23.131 &  0.008 \\
&57807.49 & 23.122 &  0.011 \\
&57818.52 & 23.129 &  0.008 \\
&57837.26 & 23.128 &  0.015 \\
&57844.33 & 23.168 &  0.038 \\
\hline
$i$&57717.61 & 23.110 &  0.027 \\
&57721.54 & 23.063 &  0.041 \\
&57747.61 & 22.961 &  0.016 \\
&57755.55 & 22.982 &  0.014 \\
&57776.53 & 22.977 &  0.009 \\
&57783.44 & 22.868 &  0.023 \\
&57786.59 & 22.939 &  0.011 \\
&57809.41 & 22.915 &  0.015 \\
&57816.48 & 22.879 &  0.010 \\
&57835.25 & 22.874 &  0.015 \\
&58195.40 & 23.147 &  0.022 \\
\hline
$z$&57715.53 & 23.363 &  0.083 \\
&57721.59 & 23.088 &  0.035 \\
&57745.56 & 22.991 &  0.023 \\
&57755.45 & 22.963 &  0.051 \\
&57774.49 & 22.871 &  0.010 \\
&57779.53 & 22.917 &  0.043 \\
&57783.55 & 22.863 &  0.014 \\
&57805.37 & 22.837 &  0.021 \\
&57816.30 & 22.854 &  0.027 \\
&57834.45 & 22.801 &  0.021 \\
&57841.40 & 22.742 &  0.029 \\
&58131.58 & 22.860 &  0.025 \\
&58159.55 & 23.102 &  0.034 \\
&58195.28 & 23.199 &  0.033 \\
\hline
$y$&57715.62 & 23.188 &  0.070 \\
&57748.54 & 23.344 &  0.216 \\
&57757.55 & 23.167 &  0.048 \\
&57778.62 & 23.301 &  0.154 \\
&57787.46 & 22.978 &  0.048 \\
&57811.40 & 23.119 &  0.046 \\
&57819.47 & 22.929 &  0.027 \\
&57833.38 & 23.042 &  0.056 \\
&57863.28 & 23.028 &  0.075 \\
\enddata
\end{deluxetable}

\begin{figure}
 \begin{center}
  \includegraphics[width=\columnwidth]{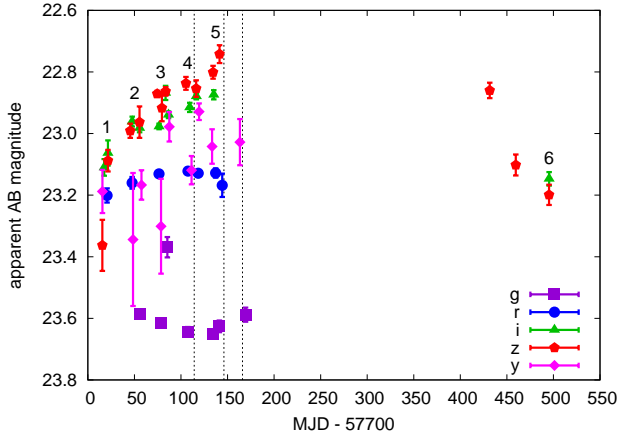}  
 \end{center}
\caption{
Observed LCs of HSC16aayt. The dashed lines show the time we performed the spectroscopic observations. The numbers in the figure correspond to the epochs shown in Fig.~\ref{fig:sed}.
}\label{fig:obslightcurve}
\end{figure}

\subsection{Photometry}
All the photometric observations reported in this paper were performed with HSC. The photometric data from Nov. 2016 to Apr. 2017 are obtained during the HSC SSP transient survey \citep{yasuda2019ssp}. In addition, one year later, the photometric data in 2018 were obtained by the Subaru Intensive Program for the transient survey in COSMOS (S17B-055I, PI: N. Suzuki).

The first-year observations in 2016-2017 were performed with all the broad-band filters of HSC, the $g$ ($4000-5450$~\AA), $r$ ($5450-7000$~\AA), $i$ ($7000-8550$~\AA), $z$ ($8550-9300$~\AA), and $y$ ($9300-10700$~\AA) bands \citep{aihara2018hscssp}. The second-year observations were performed with the $i$ and $z$ bands.

The data are reduced with hscPipe \citep{bosch2017hscpipe}, a version of the LSST stack \citep{ivezic2008lsst,axelrod2010lsst,juric2015lsst}. The astrometry and photometry are calibrated relative to the Pan-STARRS1 (PS1) $3\pi$ catalog \citep{magnier2013ps1photo,schlafly2012ps1photo,tonry2012ps1photo}. The final photometry is obtained via point-spread function photometry in the template-subtracted images. We refer to \citet{yasuda2019ssp} and \citet{aihara2018hscssp} for further details on the HSC transient survey data and the data reduction.

Fig.~\ref{fig:obslightcurve} shows the observed LC of HSC16aayt. The original data are summarized in Table~\ref{tab:photometory}. The change in the $g$ and $r$ band brightness remains within 0.1 magnitude, except for one epoch in the $g$ band. The $i$ and $z$ band brightness increases steadily during the half year survey in 2016-2017. The $i$ and $z$ band brightness is increased only by 0.2~mag and 0.6~mag, respectively, in the half year. The $y$ band photometric data are noisy but the $y$ band brightness change is also likely small during the half year.

After 290~days since the last observation in the $z$ band in 2017, HSC16aayt was observed again in 2018. The $z$ band magnitude at this epoch is similar to that at the last observation in 2017. However, the $z$ band magnitude suddenly dropped by 0.2~mag in the next 28~days and remained at a similar magnitude for the last observation. At this point, the $z$ band magnitude is similar to the discovery magnitude at around 500~days before. During the  observations in 2018, the $i$ band photometry was also obtained once and the $i$ band magnitude is similar to the $z$ band magnitude.

\begin{figure*}
 \begin{center}
  \includegraphics[width=2\columnwidth]{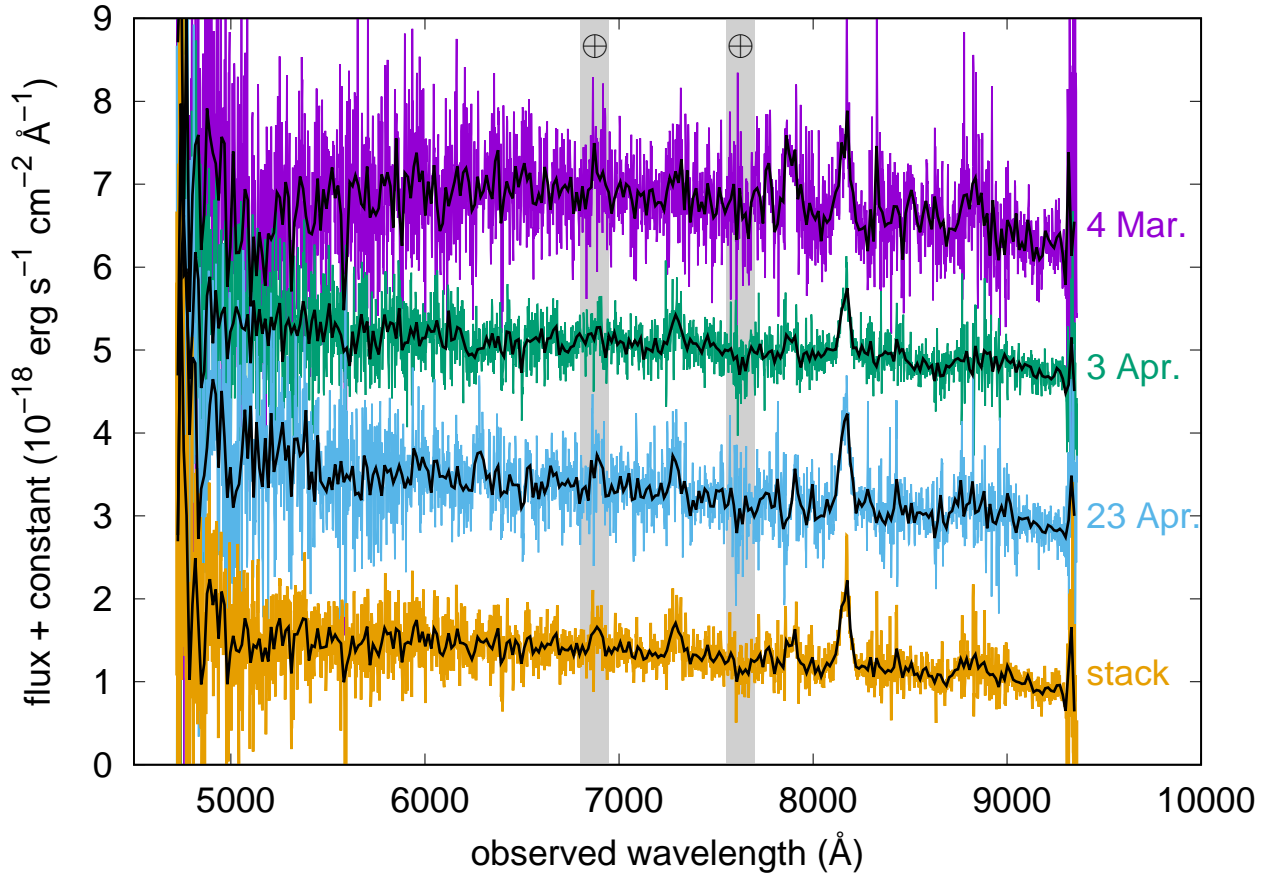}  
 \end{center}
\caption{
Observed spectra of HSC16aayt. The top three spectra were taken on the indicated dates. The bottom spectrum is obtained by stacking the three obtained spectra. The colored spectra are the original spectra and the black spectra are the binned spectra with 10 pixels. The flux is shifted by $+5\times 10^{-18}~\mathrm{erg~s^{-1}~cm^{-2}~\mbox{\AA}^{-1}}$ (4 Mar.), $+4\times 10^{-18}~\mathrm{erg~s^{-1}~cm^{-2}~\mbox{\AA}^{-1}}$ (3 Apr.), $+2\times 10^{-18}~\mathrm{erg~s^{-1}~cm^{-2}~\mbox{\AA}^{-1}}$ (23 Apr.), and $0$ (stack) for clarity. The wavelengths affected by the strong telluric absorption are shaded.
}\label{fig:spectra}
\end{figure*}

\subsection{Spectroscopy}
The spectroscopic observations were performed with Gemini/GMOS-S \citep{hook2004gmos} on 4 Mar. 2017, 3 Apr. 2017, and 23 Apr. 2017 under the program GS-2017A-Q-13\footnote{HSC16aayt was originally named as HSC16ciz and HSC16aayt is referred as HSC16ciz in the Gemini archive database.}, which was awarded through the time exchange program between Gemini and Subaru (S17A-056, PI: T. Moriya).

The instrumental configuration of the three observations was the same. The R400 grism with the GG455 filter and the 1.0\arcsec\ slit were used. This results in the spectral resolution of $R \sim 950$. A single exposure was 900~sec and 12 exposures were taken in one night, making the total exposure time in each night 3~hours. The first 6 exposures were taken by setting 7000~\AA\ as the central wavelength. The remaining 6 exposures were taken with 7100~\AA\ as the central wavelength so that the CCD chip gaps can be covered. The slit alignment was performed in the $i$ band. The first spectrum on 4 Mar. 2017 was taken with the slit angle of $21.4$~deg (from north to east) so that the host galaxy candidate at $z\sim 1.45$ was located in the slit. However, we did not find any spectroscopic signatures of the galaxy.
The slit angle was set at the average parallactic angle during the observations on 3 Apr. 2017 ($-38.94$~deg from north to east) and 23 Apr. 2017 ($-59.19$~deg from north to east). We note that the atmospheric dispersion corrector is not on GMOS-S.

The data reduction process included standard CCD processing and
spectrum extraction with IRAF\footnote{The Image Reduction and
  Analysis Facility ({\tt IRAF}) is distributed by the National
  Optical Astronomy Observatories, which are operated by the
  Association of Universities for Research in Astronomy, Inc., under
  cooperative agreement with the National Science Foundation.}. We
applied our own IDL routines to flux calibrate the data and remove
telluric lines using the well-exposed continua of the
spectrophotometric standard stars \citep{1988ApJ...324..411W,2003PASP..115.1220F}.  Details of the spectroscopic reduction are
described in \citet{2012MNRAS.425.1789S}.

The spectra of HSC16aayt are presented in Fig.~\ref{fig:spectra}. We show the three spectra taken on 4 Mar. 2017, 3 Apr. 2017, and 23 Apr. 2017. In addition, we present a spectrum in which the three spectra are stacked to improve the signal-to-noise ratio. The colored spectra in Fig.~\ref{fig:spectra} are the original spectra. The black spectra on top of the colored spectra are the binned spectra with 10 pixels. Overall, we do not find significant spectroscopic evolution in the three spectra.

We can clearly find a strong emission line at around 8150~\AA\ in all the spectra. We identify it as a hydrogen emission line because hydrogen emission lines are usually the strongest in SN spectra. The line width is relatively narrow ($\simeq 1000-2000~\kmps$, Section~\ref{sec:spectra}) and we classify HSC16aayt as a SN~IIn. In the stacked spectrum, we can further identify two possible emission lines at around 7300~\AA\ and 6900~\AA. The later two lines are, however, not clearly present in some individual spectra (Fig.~\ref{fig:spectra}). Especially, the possible emission line at 6900~\AA\ is overlapped with the strong telluric absorption.

\begin{figure*}
 \begin{center}
  \includegraphics[width=2\columnwidth]{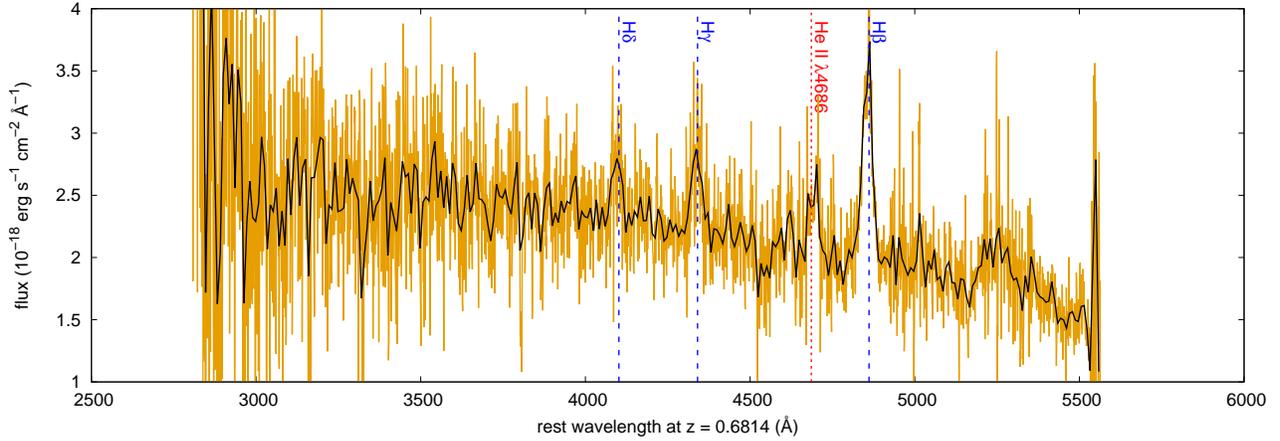} 
 \end{center}
\caption{
Stacked spectrum of HSC16aayt in the rest frame at $z=0.6814$.  The colored spectra are the original spectra and the black spectra are the binned spectra with 10 pixels. The wavelengths of \Hb, \Hg, \Hd, and He~\textsc{ii} $\lambda4686$ are shown with the vertical lines.
}\label{fig:spectraredshifted}
\end{figure*}

\begin{figure}
 \begin{center}
  \includegraphics[width=1\columnwidth]{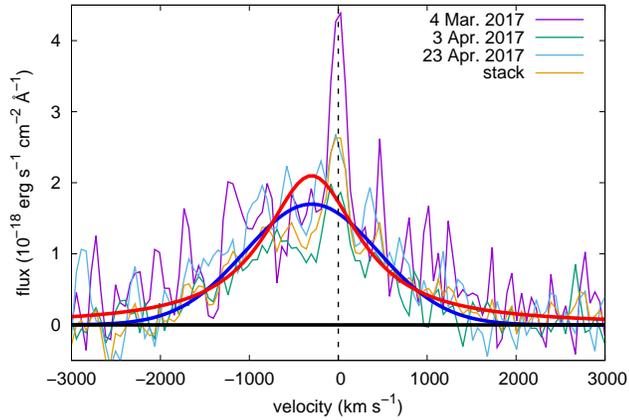}  
 \end{center}
\caption{
Continuum subtracted \Hb\ profiles in the velocity coordinate. We show the original unbinned spectra. The hydrogen emission consists of narrow and broad components. The SN redshift is determined by the peak of the narrow emission. The broad component peaks at $-300~\kmps$. It is consistent with both the Gaussian profile with the FWHM velocity of 1800~\kmps\ (blue) and the Lorentzian profile with the FWHM velocity of 1300~\kmps\ (red).
}\label{fig:h_vel}
\end{figure}

\begin{figure}
 \begin{center}
  \includegraphics[width=\columnwidth]{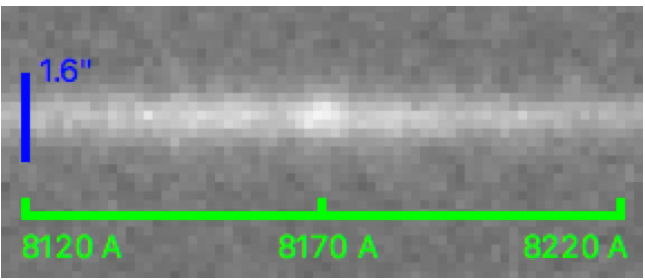}
   \end{center}
\caption{
Two dimensional spectrum of the strongest line at $8170\pm 50$~\AA\ in the observer frame from the stacked spectrum. The narrow emission component is at the center and it is on the main SN spectral component. 
}\label{fig:h_2D}
\end{figure}

\begin{figure}
 \begin{center}
  \includegraphics[width=\columnwidth]{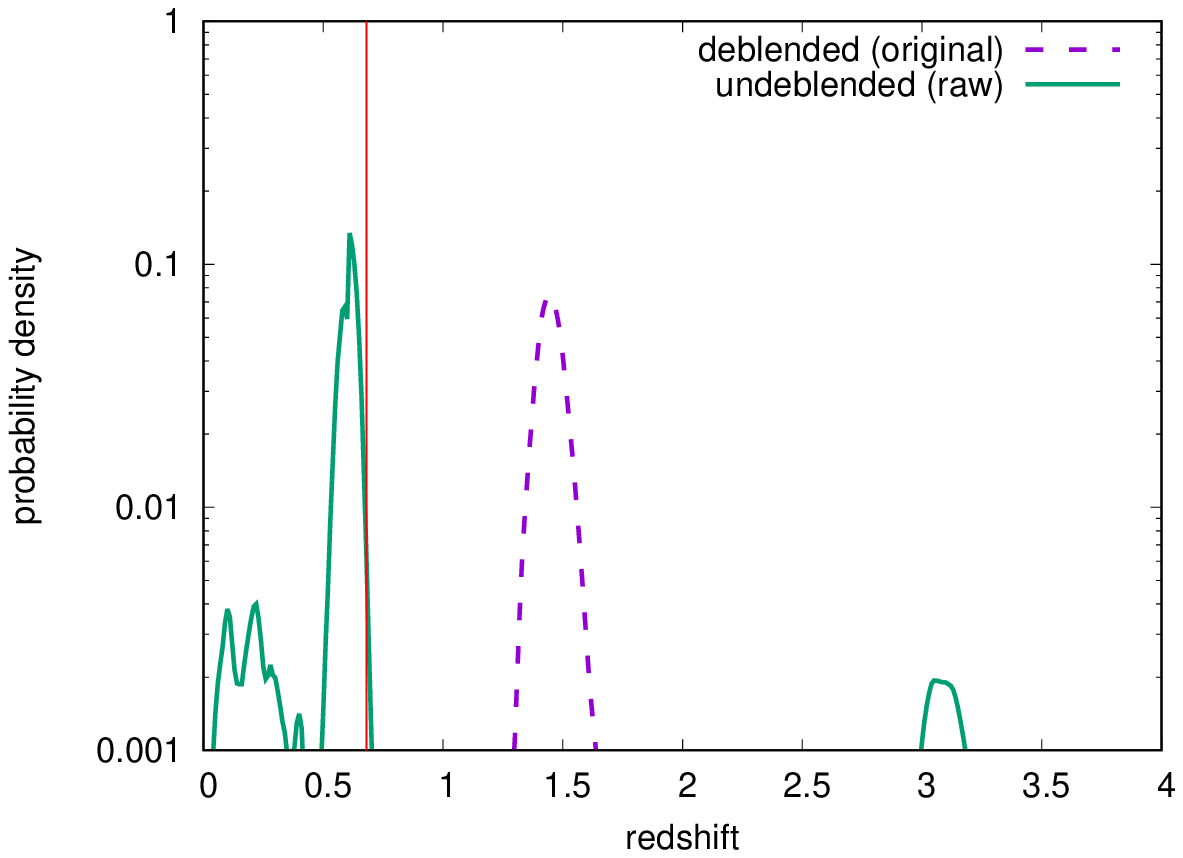} \\
  \includegraphics[width=\columnwidth]{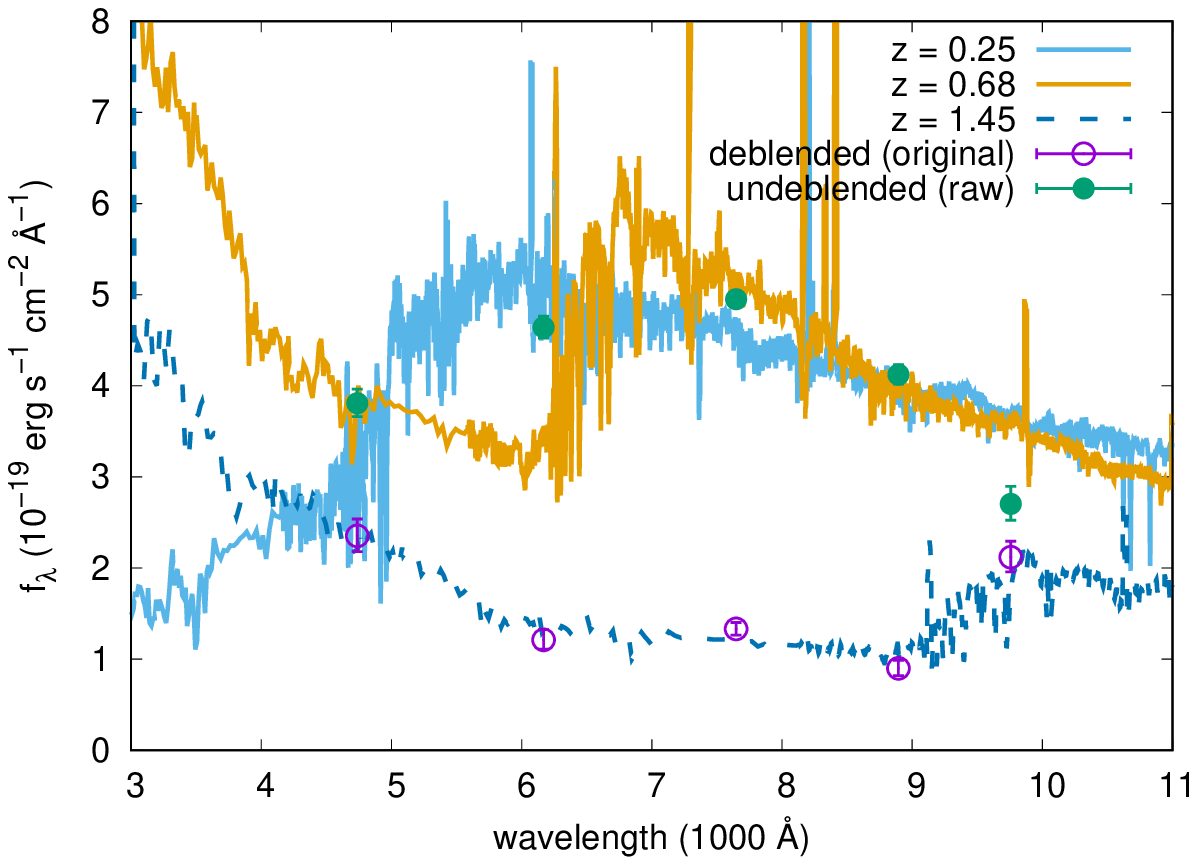}
 \end{center}
\caption{
\textit{Top:}
PDFs of the photometric redshift for the galaxy A, which is the apparent nearest galaxy to HSC16aayt on the sky (Fig.~\ref{fig:face}). We show two PDFs, one estimated from the original deblended photometry and the other from the raw undeblended photometry. The vertical line shows the redshift from the SN spectrum (0.6814).
\textit{Bottom:}
SED of the galaxy A. The best fit galaxy template at $z=1.45$ from the original deblended photometry is shown. We also show the best fit galaxy templates for the raw undeblended photometry obtained by assuming $z=0.68$ and $z=0.25$.
}\label{fig:gala}
\end{figure}

\section{Redshift and host galaxy of HSC16\MakeLowercase{aayt}}\label{sec:redshift}
We find that several emission lines can be identified when we set the SN redshift at $z=0.6814\pm 0.0002$ (Fig.~\ref{fig:spectraredshifted}). The strongest line at around 8150~\AA\ is identified as \Hb\ and the two emission lines at around 7300~\AA\ and 6900~\AA\ in the observed frame match very well to \Hg\ and \Hd, respectively. In addition, another potential emission line at around 7900~\AA\ in the observed frame matches to He~\textsc{ii} $\lambda 4686$ in the rest frame if we set the redshift to $z=0.6814$. These line identifications strongly suggest that the redshift of HSC16aayt is $z=0.6814$ and we take this redshift as the SN redshift. Other redshift possibilities are discussed in Section~\ref{sec:otherredshifts} for completeness.

Fig.~\ref{fig:h_vel} shows the continuum subtracted line profile of \Hb\ in the velocity space. It has the narrow line component on top of the broad one. Such a narrow line component is often observed in SNe~IIn. This narrow component is on top of the SN spectral component even in the two dimensional spectra (Fig.~\ref{fig:h_2D}) and it originates from the SN itself. The redshift $0.6814\pm 0.0002$ is determined by the peak of the narrow component. The redshift uncertainty came directly from the error in determining the peak of the narrow emission.

At the exact SN location, nothing is detected in our reference images with the limiting magnitudes of $\sim 27~\mathrm{mag}$ in the optical bands. Assuming that HSC16aayt is at $z \simeq 0.68$, there is nothing brighter than $\sim -16~\mathrm{mag}$ at the SN location. One possibility is that there is a underlying host galaxy at the SN location that is fainter than the limiting magnitudes in the reference images. To have a quantitative idea of the fraction of core-collapse SNe that appear in such faint galaxies at $z\simeq 0.68$, we estimate the fraction of the star formation activity occurring in galaxies fainter than $\sim -16~\mathrm{mag}$ at $z\simeq 0.68$. For this purpose, we put the star formation rate of \citet{whitaker2014galsfr} into the galaxy luminosity function of \citet{ilbert2005gallf} by assuming the mass luminosity relation of $M_\odot/L_\odot\sim 1$ to estimate the fraction of the star formation activity in the faint galaxies. We find that only  $\sim 5$\% of the star formation is from the galaxies fainter than $-16~\mathrm{mag}$ at $z\simeq 0.68$. If we simply assume that the core-collapse SN rate is proportional to the star formation rate, we expect only $\sim 5$\% of core-collapse SNe to appear in such faint galaxies. In addition, SNe~IIn are only $\sim 5$\% of core-collapse SNe \citep{shivvers2017snfrac} and, therefore, only $\sim 0.3$\% of core-collapse SNe are SNe~IIn from such faint galaxies at $z\simeq 0.68$. During our half-year transient survey, we discovered about 1400 SNe that are not clearly SNe~Ia \citep{yasuda2019ssp}. Some SNe happened to appear in galaxies with spectroscopic redshifts and they show that our SN redshift distribution is broadly populated at $0.2\lesssim z \lesssim 1.1$ \citep{yasuda2019ssp}. We find that SNe from $z\simeq 0.7\pm 0.05$ is around 10\% of all the SNe with the known redshifts. Thus, we expect $\sim 140$~SNe at $z\simeq 0.7\pm 0.05$ in our whole sample. Therefore, the expected number of hostless SNe~IIn at $z\simeq 0.7\pm 0.05$ during our half-year survey is 0.4. In addition, many thermonuclear SNe are still likely to remain in the 1400~SNe because of our strict criteria adopted to classify SNe~Ia \citep{yasuda2019ssp}. Thus, we conclude that it is unlikely to find SNe~IIn at $z\sim 0.7$ in a galaxy fainter than our reference limit during our transient survey and it is reasonable to assume that no faint host galaxy exists at the SN location. 

The nearest galaxy on the sky found around HSC16aayt is the galaxy A.
The top panel of Fig.~\ref{fig:gala} shows the probability distribution functions (PDFs) of the photometric redshift estimated for the galaxy A (Fig.~\ref{fig:face}) by the \texttt{MIZUKI} code.
Galaxy A was first considered to be a high-redshift galaxy because of the sharp peak in the original PDF at $z=1.45$.
The original apparent magnitudes used to derive this PDF are $g\simeq 25.8~\mathrm{mag}$, $r\simeq 25.9~\mathrm{mag}$, $i\simeq 25.4~\mathrm{mag}$, $z\simeq 25.5~\mathrm{mag}$, and $y\simeq 24.3~\mathrm{mag}$. The spectral energy distribution (SED) of the galaxy A from this original photometry and its best fit galaxy template at $z=1.45$ are shown in the bottom panel of Fig.~\ref{fig:gala}. Indeed, the Balmer break between 9000~\AA\ and 10000~\AA\ strongly suggested that the galaxy A is at $z\simeq 1.45$.

The photometry measurements used for the original photometric redshift estimate are, however, corrected by the deblender in the HSC pipeline \citep{bosch2017hscpipe}. This correction was made because of the extended galaxy at the north of the SN location (galaxy B in Fig.~\ref{fig:face}). We find that the images are likely oversubtracted by the deblender at the location of the galaxy A. When we take the raw apparent magnitudes before the deblending ($g\simeq 25.3~\mathrm{mag}$, $r\simeq 25.5~\mathrm{mag}$, $i\simeq 23.9~\mathrm{mag}$, $z\simeq 23.8~\mathrm{mag}$, and $y\simeq 24.1~\mathrm{mag}$), the photometric redshift becomes $0.61^{+0.04}_{-0.04}$ (Fig.~\ref{fig:gala}). The SFR of the galaxy A is esitmated to be around $0.1~\mathrm{M_\odot~yr^{-1}}$.
We note that the SN photometry is not affected by the deblending because it is measured in the subtracted images.

The redshift of HSC16aayt ($z=0.6814$) is consistent with the redshift estimate for the galaxy A when we adopt the raw undeblended photometry. If we take $z\simeq 1.45$ from the original photometric estimate, no strong lines can be identified at the wavelengths of the emission lines observed in HSC16aayt. Thus, the host galaxy of HSC16aayt is likely the galaxy A at the low redshift. We show the best fit galaxy template for the galaxy A obtained by assuming $z=0.68$ with the raw undeblended photometry in Fig.~\ref{fig:gala}. The raw undeblended photometry is found to match the template well. The SN location is 0.7\arcsec\ away from the galaxy center, which corresponds 5.0~kpc with the angular distance scale of 7.1 $\mathrm{kpc~arcsec^{-1}}$ at $z=0.6814$.

The nearby galaxies D and E on the sky (Fig.~\ref{fig:face}) also have the photometric redshifts close to $z=0.6814$ ($0.72^{+0.04}_{-0.04}$ for D and $0.76^{+0.16}_{-0.20}$ for E) and they are also host galaxy candidates of HSC16aayt at $z=0.6814$, although the galaxy A is more likely. The observed magnitudes of the galaxy D are $g\simeq 25.4~\mathrm{mag}$, $r\simeq 23.4~\mathrm{mag}$, $i\simeq 22.3~\mathrm{mag}$, $z\simeq 21.8~\mathrm{mag}$, and $y\simeq 21.4~\mathrm{mag}$, and those of the galaxy E are $g\simeq 27.2~\mathrm{mag}$, $r\gtrsim 27~\mathrm{mag}$, $i\simeq 27.4~\mathrm{mag}$, $z\simeq 27.3~\mathrm{mag}$, and $y\simeq 26.0~\mathrm{mag}$. We note that these photometric redshifts and magnitudes are based on the deblended photometry and they might also be affected by the deblending. The galaxies D and E are 5.7\arcsec\ and 3.1\arcsec\ away from the SN location, respectively, and HSC16aayt is 40~kpc (the galaxy D) or 22~kpc (the galaxy E) away from the galaxy center.

\begin{figure}
 \begin{center}
   \includegraphics[width=\columnwidth]{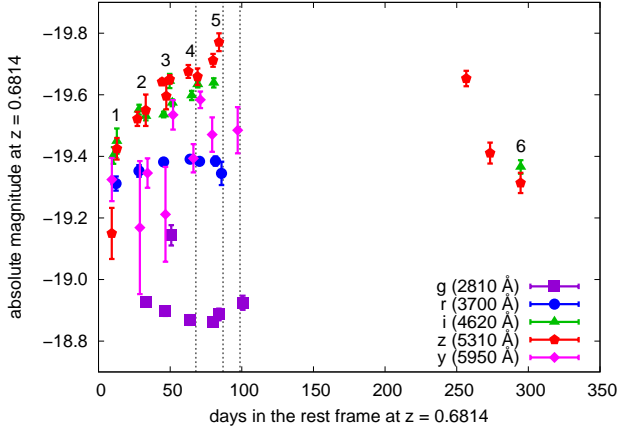}    
 \end{center}
\caption{
LCs of HSC16aayt in the rest frame at $z=0.6814$. The dashed lines show the time we performed the spectroscopic observations. The numbers in the figure correspond to the epochs shown in Fig.~\ref{fig:sed}. The simple K correction of $2.5\log (1+z)$ is applied. The observed filters and their central wavelengths at the rest frame are shown.
}\label{fig:restlightcurve}
\end{figure}

\begin{figure}
 \begin{center}
  \includegraphics[width=1\columnwidth]{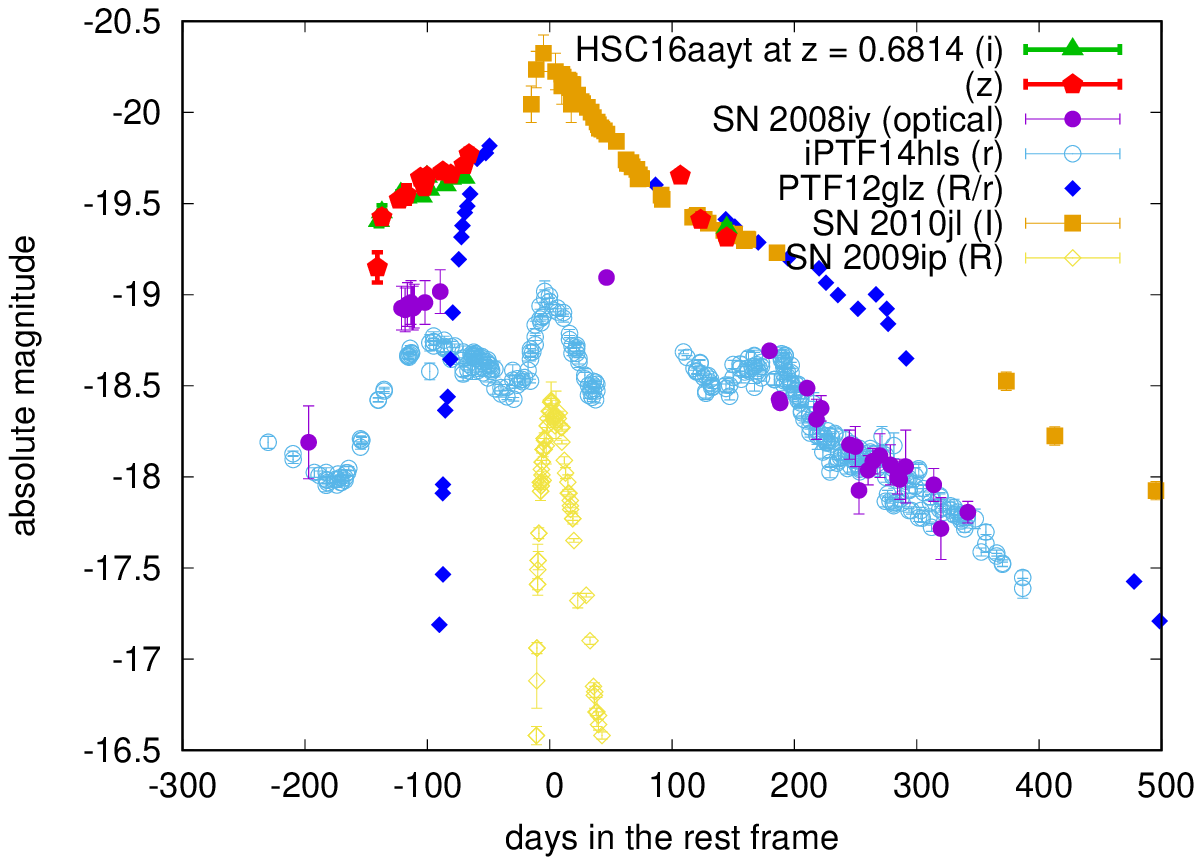}
 \end{center}
\caption{
Optical LCs of HSC16aayt and other long-timescale SNe. The observed bands are indicated in the figure. We show the LCs of SN~2008iy \citep{miller2010sn2008iy}, iPTF14hls \citep{arcavi2017iptf14hls}, PTF12glz \citep{soumagnac2018ptf12glz}, SN~2010jl \citep{fransson2014sn2010jl}, and SN~2009ip \citep{fraser2013sn09ip}.
}\label{fig:lightcurvecomparison}
\end{figure}

\begin{figure}
 \begin{center}
  \includegraphics[width=1\columnwidth]{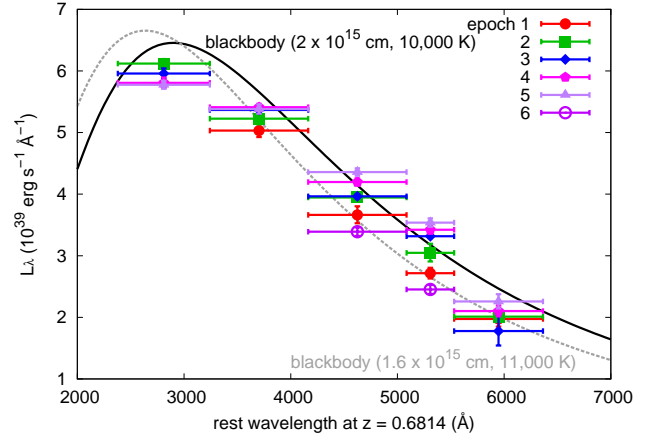}
 \end{center}
\caption{
SED of HSC16aayt based on the broad-band photometry. We also show the blackbody SEDs with two temperature and radius combinations for comparison.
}\label{fig:sed}
\end{figure}

\section{Light curve properties}\label{sec:lightcurve}
Fig.~\ref{fig:restlightcurve} shows the rest-frame LCs of HSC16aayt at $z=0.6814$. The absolute magnitudes are K corrected by using the simple correction of $2.5\log (1+z)$ \citep{hogg2002kcorr}. The observed band in the rest frame and its central wavelength at the corresponding redshift are indicated in the figure. The peak optical magnitude of HSC16aayt is likely around $-19.9~\mathrm{mag}$. This peak magnitude is common in SNe~IIn \citep[e.g.,][]{richardson2014lumfun}. The optical rise time exceeds 100~days and HSC16aayt has a very slow LC evolution.

Fig.~\ref{fig:lightcurvecomparison} compares the LC of HSC16aayt with those of other long-lasting SNe. The overall LC evolution is found to be very similar to that of SN~2008iy \citep{miller2010sn2008iy}, although SN~2008iy is about 0.5~mag fainter than HSC16aayt. SN~2008iy is a SN~IIn as is HSC16aayt, and its spectroscopic properties are also found to be similar to HSC16aayt (Section~\ref{sec:spectra}). iPTF14hls is another example of long-lasting SNe that kept its brightness as long as HSC16aayt \citep{arcavi2017iptf14hls,andrews2018iptf14hls,sollerman2019iptf14hls}. The luminosity of iPTF14hls is similar to that of SN~2008iy, but its LC has the bumpy structure that is not seen in HSC16aayt during our observation. However, we cannot exclude the possibility that the bumpy phases are missed by our observation and the sudden $z$ band brightness decrease found in 2018 in HSC16aayt is actually consistent with the brightness decrease found in iPTF14hls. The late-phase spectra of iPTF14hls, which show interaction singatures, are found to be similar to our spectra of HSC16aayt as discussed in Section~\ref{sec:spectra}.

SN~2010jl \citep[e.g.,][]{zhang2012sn2010jl,fransson2014sn2010jl,ofek2014sn2010jl} and PTF12glz \citep{soumagnac2018ptf12glz} are other examples of slowly evolving SNe~IIn and they are found to have similar luminosity to HSC16aayt. Although their LC decline rate is consistent with HSC16aayt, both SNe~IIn have much quicker rises than HSC16aayt. The major eruption of SN~2009ip in 2012 \citep[e.g.,][]{fraser2013sn09ip,prieto2013sn09ip} has much faster LC evolution than HSC16aayt (Fig.~\ref{fig:lightcurvecomparison}).

Fig.~\ref{fig:sed} presents the rest-frame SEDs of HSC16aayt which are based on the photometric data. The epochs are indicated in Figs.~\ref{fig:obslightcurve} and \ref{fig:restlightcurve}. The SEDs do not have significant evolution during our observations. Because the peak of the SEDs is not constrained by our photometric data, it is hard to determine the blackbody temperature based on the SEDs. Assuming that the bluest band corresponds to the peak of the SED, we obtain the blackbody temperatures of around $11000-10000$~K (Fig.~\ref{fig:sed}). Thus, the blackbody temperature of HSC16aayt is presumed to be kept hotter than these temperatures during our observations. The corresponding blackbody radii are $1.6\times 10^{15}$~cm (11000~K) or $2.0\times 10^{15}~\mathrm{cm}$ (10000~K). A similar small change in the blackbody temperature is observed in iPTF14hls \citep{arcavi2017iptf14hls}.

By integrating the blackbody SED with 10000~K and $2\times 10^{15}~\mathrm{cm}$, the bolometric luminosity is estimated to be $\simeq 3\times 10^{43}~\mathrm{erg~s^{-1}}$.
Because the peak of the SED is not constrained, it is possible that the actual blackbody temperature is higher and therefore the bolometric luminosity is larger. 

\begin{figure}
 \begin{center}
  \includegraphics[width=1\columnwidth]{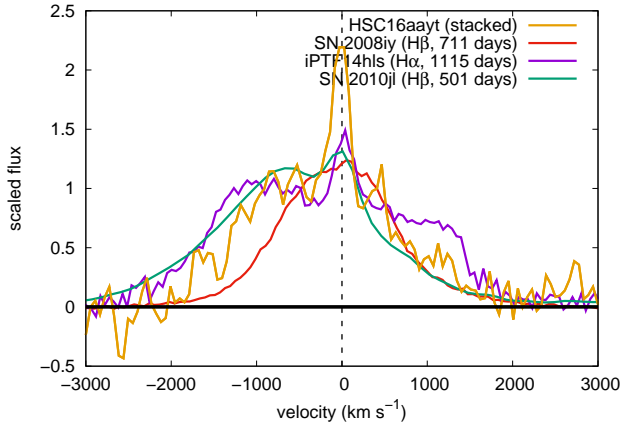}  
   \end{center}
\caption{
Comparison of the stacked spectrum of HSC16aayt and the spectra of SN~2008iy at 711~days \citep{miller2010sn2008iy}, iPTF14hls at 1115~days \citep{sollerman2019iptf14hls}, and SN~2010jl at 501~days \citep{jencson2016sn10jl}.
}\label{fig:h_vel_comp}
\end{figure}

\section{Spectroscopic properties}\label{sec:spectra}
The observed spectra of HSC16aayt are presented in Fig.~\ref{fig:spectra} and the rest-frame spectrum at $z=0.6814$ is in Fig.~\ref{fig:spectraredshifted}. Fig.~\ref{fig:h_vel} presents the strongest emission line profile (\Hb) in the velocity coordinate. 
The velocity resolution of our spectra is $\sim 300~\kmps$. The narrow emission line is identified in the spectra on 4 Mar. 2017 and 3 Apr. 2017, but it is less significant in the spectrum on 23 Apr. 2017. The narrow emission line is clearly found in the stacked spectrum and it is on the SN component as shown in Fig.~\ref{fig:h_2D}. The broad \Hb\ component can be fitted by the Gaussian profile with the full-width half-maximum (FWHM) velocity of 1800~\kmps\ or the Lorentzian profile with the FWHM velocity of 1300~\kmps\ (Fig.~\ref{fig:h_vel}). The peak of the Gaussian and the Lorentzian is both at $-300~\kmps$. This velocity shift might be caused by the radiative acceleration of the unshocked CSM (e.g., \citealt{fransson2014sn2010jl}, see also \citealt{fransson2005hst93j98s}). The velocity that can be achieved by the radiative acceleration estimated by the formula in \citet[][Eq.~1]{fransson2014sn2010jl} is indeed of the order of 100~\kmps. The \Hb\ luminosity is measured to be $\simeq 10^{41}~\mathrm{erg~s^{-1}}$. The \Ha\ luminosity to \Hb\ luminosity ratio is $\sim 3-10$ in SNe~IIn \citep[e.g.,][]{taddia2013iin,miller2010sn2008iy} and, therefore, the \Ha\ luminosity is estimated to be about 1\% of the bolometric luminosity as usually found in SNe~IIn \citep[e.g.,][]{taddia2013iin}.

Fig.~\ref{fig:h_vel_comp} compares the stacked high-signal-to-noise-ratio spectrum of HSC16aayt with the high-resolution spectrum of SN~2008iy \citep{miller2010sn2008iy} and late-phase spectra of iPTF14hls \citep{sollerman2019iptf14hls} and SN~2010jl \citep{jencson2016sn10jl}. For iPTF14hls, we show its \Ha\ emission because its \Hb\ emission was not clearly observed. The broad spectral components of these SNe are overall similar. The peaks of the broad components are shifted about $-300~\kmps$. The blue shifted \Ha\ emission is observed in iPTF14hls in the late phase although the velocity shift is larger ($\simeq -1000~\kmps$, \citealt{andrews2018iptf14hls,sollerman2019iptf14hls}). The narrow hydrogen emission similar to HSC16aayt is found in iPTF14hls as well \citep{andrews2018iptf14hls,sollerman2019iptf14hls}. We do not see a clear P-Cygni absorption component in HSC16aayt and the wind velocity around HSC16aayt is not constrained. Following the similarity between HSC16aayt and SN~2008iy, we adopt the wind velocity of 100~\kmps\ in the following discussion, which is estimated in SN~2008iy based on the narrow hydrogen P-Cygni profile.

We identify the possible emission line of He~\textsc{ii} $\lambda4686$ (Fig.~\ref{fig:spectraredshifted}). This line indicates the existence of hard photons ionizing He. This line is often observed in Wolf-Rayet stars \citep[e.g.,][]{crowther2007wrstar}. Especially, the strong He~\textsc{ii} $\lambda4686$ emission has been observed in ``flash'' spectra that are likely affected by dense CSM \citep[e.g.,][]{gal-yam2014flash,yaron2017iipcsm}. We suggest that the strong He~\textsc{ii} $\lambda4686$ emission indicates the existence of hard photons released by the CSM interaction.
We also note that such a strong He~\textsc{ii} emission ($\sim 30-50$~\% of \Hb) is sometimes found in metal-poor blue compact galaxies \citep[e.g.,][]{kehrig2018he,kehrig2011he}.

\section{Discussion}\label{sec:discussion}

\subsection{Possible isolation of HSC16aayt}
The probable host galaxy A is 5.0~kpc away from the SN location. It is difficult to determine the shape and orientation of the host galaxy A.
Because 5~kpc from the galactic center can still have star-forming activity depending on the galaxy's shape and orientation, there could be an underlying star-forming activity of the galaxy at this location. We find that about 25\% of non SNe~Ia with a known distance discovered during our HSC SSP transient survey is located beyond 5~kpc from the host galaxy center.

On the other hand, no star-forming activity is often found where SNe~IIn explode \citep[e.g.,][]{smith2016sn09iploc}. LBVs, which are progenitors of some SNe~IIn, are also suggested to be often located in remote locations (\citealt{smith2015lbviso}, but see also \citealt{humphreys2016agsmi}). In addition, SNe~II are also being discovered in remote locations from their host galaxies \citep[e.g.,][]{meza2018remoteiip}. It is also worth noting that the similar SN~IIn 2008iy exploded in a remote location from its host galaxy as well ($\simeq 8~\mathrm{kpc}$ from the center, \citealt{miller2010sn2008iy}). Therefore, it is also possible that HSC16aayt is another example of isolated SNe~IIn whose progenitor was born closer in the host galaxy, was thrown away from the birth place, and exploded in the remote location. If we assume that the progenitor was thrown away from the galaxy center, it needs to have $\simeq 500~\mathrm{km~s^{-1}}$ to travel 5~kpc assuming a typical massive star lifetime of 10~Myr. Thus, the progenitor may have been a hypervelocity star. The velocity of around 500~\kmps\ is often observed in Galactic hypervelocity stars \citep[e.g.,][]{palladino2014hypergk,brown2018gaiahv}. The origin of such a large velocity could be the SN kick \citep[e.g.,][]{tauris2015bineje} or the supermassive black hole (BH) at the host galaxy center \citep[e.g.,][]{hills1988hvs,yu2003hvs}. Although we assumed 10~Myr as a typical lifetime of massive stars, it is also possible that the progenitor had a significantly longer lifetime than 10~Myr due to, e.g., binary interaction \citep[e.g.,][]{smith2015lbviso,zapartas2017delayedcc}. The progenitor may have even been originally in a hypervelocity binary system \citep[e.g.,][]{lu2007hvb,sesana2009hvb,wang2018hvb} and eventually merged to have a longer lifetime than a single massive hypervelocity star. Then, the required velocity is lower. 

Although the galaxy A is likely the host galaxy of HSC16aayt at $z=0.6814$, there are two more galaxies near HSC16aayt that have the photometric redshifts of $z\sim 0.68$ (Section~\ref{sec:redshift}, Fig.~\ref{fig:face}): the galaxies D (photometric redshift of $0.72^{+0.04}_{-0.04}$) and E (photometric redshift of $0.76^{+0.16}_{-0.20}$). The galaxy centers are 40~kpc (D) or 22~kpc (E) away from the SN location assuming $z=0.6814$. If they were the host galaxy instead of the galaxy A, the progenitor should have been moving with the tangential velocity of 4000~\kmps\ (D) or 2200~\kmps\ (E) assuming a typical massive star lifetime of 10~Myr. No hypervelocity stars are known to have such high velocities \citep{brown2015hypervel}. However, we note that the progenitor of SN~1997C may have had a velocity larger than $\simeq 3000~\kmps$ \citep{zinn2011hostless}. \citet{zinn2011hostless} also reports several SNe whose progenitor may have had velocities close to 1000~\kmps.

\begin{figure}
 \begin{center}
  \includegraphics[width=\columnwidth]{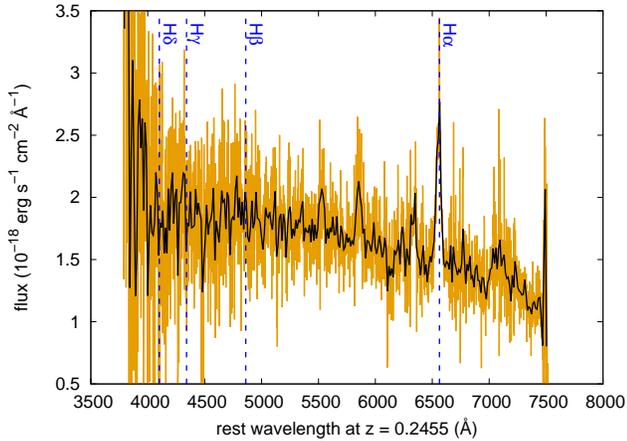}  
 \end{center}
\caption{
Stacked rest-frame spectrum of HSC16aayt if it is at $z=0.2455$.  The colored spectra are the original spectra and the black spectra are the binned spectra with 10 pixels. The wavelengths of \Ha, \Hb, \Hg, and \Hd\ are shown with the vertical lines.
}\label{fig:spectra_at_z0p2455}
\end{figure}

\subsection{Other redshift possibilities}\label{sec:otherredshifts}
We set the redshift of HSC16aayt as $z=0.6814$ because several emission lines can be identified with this redshift. However, the weak emission lines at around 7300~\AA\ and 6900~\AA\ are marginally detected in some spectra and one may doubt their significance. In this case, we need to rely only on the strongest emission line at around 8150~\AA\ to determine the redshift of HSC16aayt. Although we believe that HSC16aayt is very likely at $z=0.6814$, we discuss other redshift possibilities for completeness in this section. Determination of redshifts and host galaxies for transients without an apparent host galaxy at the SN location would be an important issue in the era of excessive transient discovery with, e.g., the Large Synoptic Survey Telescope\footnote{\url{https://www.lsst.org/}} \citep[e.g.,][]{gupta2016hostid}.

\subsubsection{$z=0.2455$}
If we identify the strongest line at around 8150~\AA\ is \Ha, HSC16aayt is at $z=0.2455\pm0.0002$. \Ha\ is usually the strongest line in hydrogen-rich interacting SNe. However, we do not see any other lines of the Balmer series in this case (Fig.~\ref{fig:spectra_at_z0p2455}). No strong emission is expected at the wavelengths of the other possible emission lines. The peak optical magnitudes of HSC16aayt at $z=0.2455$ become $\simeq -17.7~\mathrm{mag}$ and they are in the range of SNe~IIn \citep{richardson2014lumfun}.

If we take $z=0.2455$, our reference images with the limiting magnitudes of $\sim 27~\mathrm{mag}$ indicate that no galaxy brighter than $\sim -13~\mathrm{mag}$ in optical exists at the exact SN location. It is not likely to find SNe~IIn in such a faint galaxy during our survey (Section~\ref{sec:redshift}) and we argue that no host galaxy exists at the SN location even if we take this redshift possibility. The nearest galaxy on the sky (the galaxy A in Fig.~\ref{fig:face}) could be at $z\sim 0.25$ (Fig.~\ref{fig:gala}). With the $3.86~\mathrm{kpc~arcsec^{-1}}$ at $z=0.2455$, the 0.7\arcsec\ separation corresponds only to the distance of 2.7~kpc from the center of the galaxy A.
An interesting coincidence is that the extended galaxy at the north of HSC16aayt (the galaxy B in Fig.~\ref{fig:face}) is at $z=0.25053\pm0.00005$, which is spectroscopically measured by the Sloan Digital Sky Survey\footnote{\url{https://www.sdss.org/}} (SDSS). Therefore, this galaxy B can also be the host galaxy of HSC16aayt if it is at $z=0.2455$, although the galaxy center at (RA, Dec) = (10:02:05.46, +02:58:10.62) has a separation of 12.4\arcsec\ from HSC16aayt. The angular scale at $z=0.2505$ is 3.91~$\mathrm{kpc~arcsec^{-1}}$ and the physical distance between the galaxy center to the SN location is 48.5~kpc at $z=0.2505$. Again, assuming a typical massive star lifetime of 10~Myr, the progenitor needs to have the tangential velocity of 4700~\kmps\ to travel 48.5~kpc before the explosion. Another intriguing fact is the redshift difference between the host galaxy ($z=0.2505$) and HSC16aayt ($z=0.2455$). Assuming that HSC16aayt is actually at $z=0.2505$, the redshift difference indicates that HSC16aayt is moving away from the host towards us with the velocity of 1500~\kmps. Combined with the tangential velocity of 4700~\kmps, the progenitor of HSC16aayt was moving away from the host galaxy with a velocity of 4900~\kmps\ if the galaxy B is the host galaxy. 

The SDSS spectrum of the galaxy B shows that it is an active galactic nucleus. Thus, we can estimate its central BH mass from its spectrum in this case. By using the \Ha\ luminosity ($3.6\times 10^{41}~\mathrm{erg~s^{-1}}$) and the FWHM of the \Ha\ emission from the host galaxy (46.8~\AA\ in the rest frame) obtained by the SDSS spectrum, the central BH mass is estimated to be $9\times 10^6~\Msun$ \citep{greene2007bhmass}. The velocity that can be gained by the BH interaction through the Hills mechanism is
\begin{equation}
    3400 \left( \frac{a}{0.1~\mathrm{AU}} \right)^{-1/2}\left(\frac{M_\mathrm{total}}{10~\Msun} \right)^{1/3}\left(\frac{M_\mathrm{BH}}{9\times10^{6}~\Msun} \right)^{1/6}~f_R~\kmps,
\end{equation}
where $a$ is the binary separation, $M_\mathrm{total}$ is the total mass of the binary system, $M_\mathrm{BH}$ is the BH mass, and $f_R$ is a factor of unity \citep{brown2015hypervel}. The extreme velocities of the order of 1000~\kmps\ estimated above can be achieved by the BH at the center of the galaxy B.

\subsubsection{$z=1.92$}
This redshift possibility appears if we assume the strongest line is Mg~\textsc{ii} at 2800~\AA\ in the rest frame. It is one of the strongest lines in ultraviolet in SNe~IIn \citep[e.g.,][]{fransson2002sn95n,fransson2014sn2010jl}. However, no other possible emission lines can be identified in this redshift. If HSC16aayt is at $z=1.92$, the peak SN magnitudes in the rest-frame ultraviolet are around $-22~\mathrm{mag}$ and HSC16aayt is a long-lasting SLSN~IIn like SN~2006gy \citep{smith2010sn06gy}. 

If we assume $z=1.92$, there can exist a host galaxy fainter than $-17.7~\mathrm{mag}$ at the SN location and we cannot exclude the possibility that there is a non-detected host galaxy at the SN location. The galaxy A is again not likely to be a host galaxy at $z=1.92$ (Fig.~\ref{fig:gala}). The nearest galaxy on the sky in our image having a photometric redshift that is consistent with $z=1.92$ is the galaxies G (the photometric redshift of $1.4^{+0.48}_{-0.08}$) and J (the photometric redshift of $2.04^{+0.04}_{-0.04}$). The galaxy G is 8.0\arcsec\ away from the SN location and the galaxy J is 9.1\arcsec\ away. The corresponding physical distances at $z=1.92$ (8.4 kpc~$\mathrm{arcsec}^{-1}$) are 67~kpc (galaxy G) and 76~kpc (galaxy J).

\section{Summary}\label{sec:summary}
We have reported our discovery of SN~IIn HSC16aayt. It was discovered in the COSMOS field during the HSC SSP transient survey conducted from Nov. 2016. The $i$ and $z$ band magnitudes continue to brighten during the half year survey. The follow-up observations in early 2018 showed that it faded and was at around the discovery magnitudes. Therefore, the rise time of HSC16aayt is more than 100~days. 
Our spectroscopic follow-up observations revealed that HSC16aayt is a SN~IIn. The SN spectra indicate that the SN is at $z=0.6814\pm0.0002$. The peak absolute magnitudes in the $i$ and $z$ bands are estimated to be around $-19.9~\mathrm{mag}$. Overall properties of HSC16aayt are similar to those of SN~2008iy, but HSC16aayt is more luminous.


The host galaxy center is 5~kpc away from the SN location and there could be an extended star-formation activity of this galaxy at the SN location. Alternatively, the massive progenitor may have thrown away from the host galaxy by the SN kick or the interaction with the supermassive BH at the galaxy center to explode at an isolated location.

\acknowledgments
This research is supported by the Grants-in-Aid for Scientific Research of the Japan Society for the Promotion of Science (JP16H07413, JP17H02864, JP18K13585) and by JSPS Open Partnership Bilateral Joint Research Project between Japan and Chile.
Support for G.P. is provided by the Ministry of Economy, Development, and Tourism's Millennium Science Initiative through grant IC120009, awarded to The Millennium Institute of Astrophysics, MAS

This work made use of the Open Supernova Catalog (\url{https://sne.space/}, \citealt{guillochon2017osc}) and WISeREP  (\url{https://wiserep.weizmann.ac.il}, \citealt{yaron2012wiserep}).

The Hyper Suprime-Cam (HSC) collaboration includes the astronomical communities of Japan and Taiwan, and Princeton University.  The HSC instrumentation and software were developed by the National Astronomical Observatory of Japan (NAOJ), the Kavli Institute for the Physics and Mathematics of the Universe (Kavli IPMU), the University of Tokyo, the High Energy Accelerator Research Organization (KEK), the Academia Sinica Institute for Astronomy and Astrophysics in Taiwan (ASIAA), and Princeton University.  Funding was contributed by the FIRST program from Japanese Cabinet Office, the Ministry of Education, Culture, Sports, Science and Technology (MEXT), the Japan Society for the Promotion of Science (JSPS), Japan Science and Technology Agency (JST), the Toray Science Foundation, NAOJ, Kavli IPMU, KEK, ASIAA, and Princeton University.

The Pan-STARRS1 Surveys (PS1) have been made possible through contributions of the Institute for Astronomy, the University of Hawaii, the Pan-STARRS Project Office, the Max-Planck Society and its participating institutes, the Max Planck Institute for Astronomy, Heidelberg and the Max Planck Institute for Extraterrestrial Physics, Garching, The Johns Hopkins University, Durham University, the University of Edinburgh, Queen's University Belfast, the Harvard-Smithsonian Center for Astrophysics, the Las Cumbres Observatory Global Telescope Network Incorporated, the National Central University of Taiwan, the Space Telescope Science Institute, the National Aeronautics and Space Administration under Grant No. NNX08AR22G issued through the Planetary Science Division of the NASA Science Mission Directorate, the National Science Foundation under Grant No. AST-1238877, the University of Maryland, and Eotvos Lorand University (ELTE).

This paper makes use of software developed for the Large Synoptic Survey Telescope. We thank the LSST Project for making their code available as free software at \url{http://dm.lsst.org}.

Based in part on data collected at the Subaru Telescope and retrieved from the HSC data archive system, which is operated by the Subaru Telescope and Astronomy Data Center at National Astronomical Observatory of Japan.

Based in part on data obtained at the Gemini Observatory via the time exchange program between Gemini and the Subaru Telescope processed using the Gemini IRAF package (program ID: S17A-056, GS-2017A-Q-13). The Gemini Observatory is operated by the Association of Universities for Research in Astronomy, Inc., under a cooperative agreement with the NSF on behalf of the Gemini partnership: the National Science Foundation (United States), the National Research Council (Canada), CONICYT (Chile), Ministerio de Ciencia, Tecnolog\'{i}a e Innovaci\'{o}n Productiva (Argentina), and Minist\'{e}rio da Ci\^{e}ncia, Tecnologia e Inova\c{c}\~{a}o (Brazil).

Numerical analyses were in part carried out on analysis servers at Center for Computational Astrophysics, National Astronomical Observatory of Japan.

\facilities{Subaru/HSC, Gemini/GMOS-S} 
\software{MIZUKI \citep{tanaka2015mizuki}, hscPipe \citep{bosch2017hscpipe}}

\bibliographystyle{yahapj}
\bibliography{references}


\end{document}